\documentclass{ifacconf}

\usepackage{graphicx}      
\usepackage{natbib}        
\usepackage{amsmath}
\begin{document}
\begin{frontmatter}

\title{The Missing Variable: Socio-Technical Alignment in Risk Evaluation}


\author[First]{Niclas Flehmig}, 
\author[Second]{Mary Ann Lundteigen}, 
\author[First]{Shen Yin}

\address[First]{Department of Mechanical and Industrial Engineering, Norwegian University of Science and Technology (NTNU), 
   Trondheim, Norway}
\address[Second]{Department of Engineering Cybernetics, Norwegian University of Science and Technology (NTNU), Trondheim, Norway}

\begin{abstract}                
This paper addresses a critical gap in the risk assessment of AI-enabled safety-critical systems. While these systems, where AI systems assist human operators, function as complex socio-technical systems, existing risk evaluation methods fail to account for the associated complex interaction between human, technical, and organizational components. Through a comparative analysis of system attributes from both socio-technical and AI-enabled systems and a review of current risk evaluation methods, we confirm the absence of explicit socio-technical considerations in standard risk expressions. To bridge this gap, we introduce a novel socio-technical alignment ($STA$) variable designed to be integrated into the traditional risk equation. This variable estimates the degree of harmonious interaction between the AI systems, human operators, and organizational processes. A case study on an AI-enabled liquid hydrogen ($LH_2$) bunkering system demonstrates the variable's relevance. By comparing a naive and a safeguarded system design, we illustrate how the $STA$-augmented expression captures socio-technical safety implications that traditional risk evaluation overlooks, providing a more system-theoretic basis for risk evaluation.
\end{abstract}


\end{frontmatter}

\section{Introduction}
The integration of AI systems as decision-support tools for human operators in safety-critical systems holds potential for enhancing overall system safety. For instance, by predicting the dispersion of hazardous gases, an AI system can provide operators with advance warning, enabling timely intervention to prevent severe accidents or avoid costly emergency shutdowns. This diffusion of AI systems, however, compels regulators to require rigorous risk assessment and management throughout the system lifecycle, as outlined in emerging standards and legislation \citep{euaiactEuropeanUnionArtificial2024, iso/iectr5469:2024ArtificialIntelligenceFunctional2024}. While researchers increasingly recognize AI-enabled systems as complex systems which require system-theoretic approaches \citep{dobbeSystemSafetyArtificial2022a, lootzRiskManagementUncertainty2025}, this understanding is not yet fully incorporated in established risk evaluation practices. Although methods like system-theoretic process analysis (STPA) are being adopted for risk identification in complex systems \citep{koesslerRiskAssessmentAGI2023}, the socio-technical attributes which are complex interaction between human, technical, and organizational elements are not explicitly represented in existing risk equations. This work addresses the missing consideration of socio-technical attributes in the risk expression by introducing a novel socio-technical variable. Focusing on AI-enabled systems where AI systems assist human operators, we analyze the current methods for risk evaluation and propose extending the foundational risk equation with a novel socio-technical alignment variable. This work seeks to formally incorporate the critical influence of interaction between human, technical, and organizational elements into the evaluation of risk for AI-enabled systems.

\section{Definitions}\label{sec:2}

\subsection{Risk} The traditional definition of risk $R$ goes back to \cite{kaplanQuantitativeDefinitionRisk1981} and can be expressed by a set of triplets:
\begin{equation} \label{eq:risk}
R = \{<s_i,f_i,c_i>\}_{i=1}^{n}
\end{equation}
where $s_i$ describes the accident scenario, $f_i$ estimates the likelihood of an accident scenario, and $c_i$ is multidimensional and represents the potential harms and their associated probabilities. This equation can be used during the risk assessment process to evaluate the risk.

\subsection{Socio-technical system} We define socio-technical systems based on \cite{rasmussenRiskManagementDynamic1997} and \cite{zareiSafetyCausationAnalysis2024} as systems that are characterized by the complex, dynamic interdependence between its social and technical elements. More than just hardware and software, such systems integrate human operators, organizational structures, work processes, and the external environment. Their defining feature is the critical relationship between human factors, technology, and organizational context, all of which interact to achieve specific objectives.

\subsection{AI system} Based on the \cite{euaiactEuropeanUnionArtificial2024}, we define an AI system as a machine-based system designed with varying degrees of autonomy to perform tasks that traditionally require human intelligence, such as learning, prediction, recommendation, problem-solving, natural language processing, and decision-making. For instance, an end-to-end pipeline that first preprocesses an input image, then feeds it to a deep learning model that classifies the image according to the objective, and returns a probability score for the different classes.

\subsection{AI-enabled safety-critical system} An AI-enabled system can be described as a system that incorporates one or more AI systems \citep{detnorskeveritasAssuranceAIenabledSystems2023}. The rest of the underlying system can consist of components such as human operators, electronic components and non-AI software. This definition can be extended to AI-enabled safety-critical systems where a failure of the system can cause harm to people and the environment. For instance, a $LH_2$ bunkering system where an AI system is responsible for predicting the dispersion of $LH_2$ and for sending a warning to the bunkering control room. In this work, these two terms will be used interchangeably.

\section{Research approach}\label{sec:3}
The primary objective of this work is to address the emerging challenge of evaluating risk in AI-enabled systems and to answer the following three research questions:
\begin{itemize}
    \item \textbf{RQ1}: \textit{How does the integration of AI-enabled systems transform the key attributes of a dynamic socio-technical system?}
    \item \textbf{RQ2}: \textit{How do common methods for evaluating risk in AI-enabled systems express risk in the context of socio-technical safety implications?}
    \item \textbf{RQ3}: \textit{How can we determine and incorporate socio-technical safety implications into the risk evaluation of AI-enabled systems?}
\end{itemize}
To answer these questions, the research was conducted in four steps:
\begin{enumerate}
    \item Compare key attributes of socio-technical system with AI-enabled system;
    \item Review risk evaluation methods for traditional and AI-enabled systems in the context of socio-technical systems;
    \item Derive a socio-technical alignment variable for the risk expression of AI-enabled systems;
    \item Apply the socio-technical alignment variable in the risk evaluation of a case study and compare it to the traditional method.
\end{enumerate}

\subsection{Key attributes for socio-technical systems}
The work by \cite{zareiSafetyCausationAnalysis2024} outlines the core tenets of \textit{socio-technical systems theory} and its key principles and characteristics. We chose this textbook to identify the key attributes of such systems which are: \textit{interdependence}, \textit{complexity}, \textit{emergence}, \textit{non-linearity}, and \textit{adaptability}. Then, we evaluated whether AI-enabled systems can also be classified as socio-technical by comparing these attributes with those of AI-enabled systems, as informed by existing literature \citep{dobbeSystemSafetyArtificial2022a, hendrycksIntroductionAISafety2024a} and our own analysis.

\subsection{Current risk evaluation methods}
Next, we performed a literature review on \textit{risk assessment} and \textit{risk management} for both traditional and AI-enabled systems. The search covered publications from \textit{Web of Science}, \textit{Google Scholar}, and \textit{arXiv}, and used keyword combinations such as “AI risk evaluation”, “AI safety assessment”, “socio-technical safety”, and “AI-enabled systems”. Only peer-reviewed papers, standards, and authoritative reports were included. The retrieved sources were screened for whether their risk expressions explicitly represent socio-technical attributes in addition to probability- or consequence-based terms. Across all reviewed works, no method was found to incorporate such socio-technical considerations within the mathematical risk expression.\\
Our investigation began with an overview linking techniques from different safety-critical industries to AI-enabled systems \citep{koesslerRiskAssessmentAGI2023} and foundational standards for traditional risk management from the International Organization for Standardization (ISO) and the International Electrotechnical Commission (IEC) \citep{RiskManagementGuidelines2018, iec310102019Risk2020}. This was supplemented by a comprehensive textbook on risk assessment \citep{rausandRiskAssessmentTheory2020}. For AI system specific approaches, we consulted the AI risk management standard from ISO/IEC \citep{InformationTechnologyArtificial2024, AI_managementSystem}, a review from the Norwegian Ocean Safety Authority \citep{lootzRiskManagementUncertainty2025}, and recent research contributions \citep{novelliAIRiskAssessment2024, camposFrontierAIRisk2025, nagbolDesigningRiskAssessment2021}. Finally, we included a key textbook on AI safety \citep{hendrycksIntroductionAISafety2024a} for its relevant insights on risk evaluation.

\subsection{Derivation of socio-technical alignment variable}
Then, we used the results from the previous step and extended the traditional risk expression from eq.~(\ref{eq:risk}) to incorporate a socio-technical dimension. This derivation was informed by a review of relevant literature \citep{zareiSafetyCausationAnalysis2024, levesonEngineeringSaferWorld2012, rausandRiskAssessmentTheory2020}. To deconstruct the components of the standard risk expression, we first consulted foundational texts on traditional risk assessment techniques \citep{rausandRiskAssessmentTheory2020}. This approach aligns with recommendations to adapt established best practices from other industries for AI-enabled systems \citep{koesslerRiskAssessmentAGI2023}. To formulate a socio-technical variable, we then examined work that establish the intrinsic link between socio-technical systems and safety \citep{zareiSafetyCausationAnalysis2024, levesonEngineeringSaferWorld2012} as well as work on AI safety that advocates for a socio-technical framing of AI-enabled systems \citep{dobbeSystemSafetyArtificial2022a, rajiConcreteProblemsAI2024}. This link is relevant, as safety is formally defined as the freedom from risk \citep{iso/iectr5469:2024ArtificialIntelligenceFunctional2024}.

\subsection{Design criteria for relevant case study}
Finally, to design a qualitative case study demonstrating the relevance of the introduced socio-technical alignment variable within the risk expression, we applied two core criteria. First, to ensure the demonstration is grounded in an appropriate context, the selected system must itself be an AI-enabled system that is characterized by its dynamic complexity and the deep interdependence of its technical and social components. Second, to illustrate the variable's specific contribution, its relevance must be shown through a qualitative comparison. We therefore selected a single AI-enabled system with two distinct system designs. First, a design with a naive AI system incorporation, and then a safeguarded system design based on \cite{flehmigImplementingArtificialIntelligence2024a}. The safeguarded system design introduces two new components that aim to reduce the risk stemming from the AI system. To compare the risk equations, it is not the risk of the individual systems that is evaluated, but rather how the incorporation of two new components affects the risk equations. Thereby, giving a perspective on potential shortcomings of the traditional risk equation capturing changes in socio-technical attributes and showcasing the usefulness of a socio-technical alignment variable.

\section{Comparison of socio-technical systems and AI-enabled systems}\label{sec:4}
In this section, we represent the results of our comparison between socio-technical key attributes and AI-enabled systems. As defined in Section~\ref{sec:2}, a socio-technical system is a complex and dynamic system that represents the interactions between human, technical, and organizational elements. It is characterized by five different key attributes \citep{zareiSafetyCausationAnalysis2024}:
\begin{itemize}
    \item \textbf{Interdependence}: Denotes that system behavior is generated by the interactions between social and technical components. For instance, in an AI-enabled system, the AI system alert directly triggers a human operator's action in a control room.
    \item \textbf{Complexity}: Refers to the state of having many interconnected components with feedback loops, a condition arising from integrating technology, human behavior, and organizational processes. For instance, in an AI-enabled system, the system involves an AI system, operator cognition, team protocols, and regulations, creating feedback loops.
    \item \textbf{Emergence}: The phenomenon where novel, unanticipated system behaviors arise from component interactions, preventing the prediction of certain accidents from isolated analysis. Complex human-technology-organization interactions are a typical contributing factor. For instance, in an AI-enabled system, an organizational culture of automation complacency can unexpectedly arise from the combination of a highly reliable AI system and untrained, inattentive operators.
    \item \textbf{Non-linearity}: Indicates that cause-and-effect relationships are not proportional. Minor perturbations can lead to major consequences, often due to synergistic effects exceeding the sum of individual impacts. For instance, in an AI-enabled system, a minor sensor glitch that affects the AI system's output, combined with slight operator doubt, can synergistically lead to a critical warning being ignored.
    \item \textbf{Adaptability}: The system's inherent capability to adjust in response to changes in its environment, technology, and organizational needs. For instance, in an AI-enabled system, the system can reconfigure AI system parameters and retrain operators in response to new regulations or technology.
\end{itemize}
Based on the preceding attributes, AI-enabled systems that function as decision-support tools for human operators can be classified as complex socio-technical systems. This classification stems from their inherent integration of human operators, various technical components, including one or more AI systems, and the organizational processes of development and deployment. Consequently, the system's design and the nature of the interactions between human operators, AI and non-AI technology, and organizational processes are critical determinants of safe operation. Some safety implications can be:
\begin{itemize}
    \item \textbf{Organizational culture}: A robust and open safety culture that prioritizes safety, transparent communication, and incident-based learning. This is crucial for integrating AI-enabled systems effectively and enhancing overall system safety.
    \item \textbf{Human error}: Safety is impacted by human error, including decision-making mistakes and procedural violations. Mitigating these risks requires strategies like human factors training, crew resource management, and formal error management.
    \item \textbf{Human-AI interface}: The design of human–AI interfaces, including output displays and monitoring systems, must account for cognitive limitations and human factors to prevent information overload and maintain operator situational awareness.
\end{itemize}
Accident analysis in an AI-enabled system necessitates examining the interactions among human operators, technical components including the AI systems themselves, and the organizational structure that deployed them.

\section{Socio-technical attributes in the risk evaluation of AI-enabled systems}\label{sec:5}
We find that current risk evaluation methods for traditional and AI-enabled systems lack explicit consideration of socio-technical attributes. The standard risk equation Equation~\ref{eq:risk} and supporting documents \citep{RiskManagementGuidelines2018, iec310102019Risk2020, rausandRiskAssessmentTheory2020} do not define a variable for socio-technical attributes. Even literature advocating for system-theoretic approaches to AI system risk \citep{koesslerRiskAssessmentAGI2023} focuses solely on identification, namely system-theoretic process analysis (STPA), not on evaluation or its mathematical formulation. This gap persists in the latest AI system risk management guidance \cite{InformationTechnologyArtificial2024}, which offers no expression for evaluating socio-technical safety attributes.\\
\cite{hendrycksIntroductionAISafety2024a} modifies the classical risk equation for AI systems by pairing probability $P$ and hazard severity. To reflect human contextual factors, the author further augments the expression with parameters for \textit{exposure} and \textit{vulnerability}. This can be expressed by:
\begin{equation}
    \resizebox{.9\hsize}{!}{$\text{Risk} = \sum_{h} P(h) \times \text{severity}(h) \times \text{exposure}(h) \times \text{vulnerability}(h)$}
\end{equation}
where $h$ indexes the different hazardous events, $P(h)$ denotes the probability of each hazardous event, and the product of $severity(h)$, $exposure(h)$, and $vulnerability(h)$ captures how severe, frequent, and context-dependent each event is. While \cite{hendrycksIntroductionAISafety2024a} acknowledges the complexity of AI systems and the importance of a system-theoretic perspective, the proposed risk equation does not explicitly capture socio-technical attributes. Nevertheless, we include this formulation in our work for two reasons: its derivation for AI systems, and the authoritative standing of the textbook in the field of AI safety.
The proposed risk assessments for AI-enabled systems also fail to incorporate socio-technical safety implications. \cite{nagbolDesigningRiskAssessment2021} address human-AI interactions through interpretability, interface, and user competence, yet their approach omits the necessary organizational context and a formal risk expression. \cite{novelliAIRiskAssessment2024} define risk through the interaction of severity and likelihood, derived from factors like hazards, exposure, and vulnerability, but still do not account for socio-technical dimensions of the system. Finally, \cite{camposFrontierAIRisk2025} propose evaluation via key risk and control indicators, which similarly do not consider socio-technical safety attributes.

\section{Socio-technical alignment state for AI-enabled systems}\label{sec:6}
In the following, we introduce a variable that represents the socio-technical attributes and their alignment in an AI-enabled system. We define the variable $STA$, an estimate representing the degree of harmonious interaction and shared objective function among the core subsystems of an AI-enabled socio-technical system: AI system, human operator, and organizational processes. We extend the Equation~\ref{eq:risk} and introduce the new variable:
\begin{equation} \label{eq:new_risk}
{R = STA \times \{<s_i,f_i,c_i>\}_{i=1}^{n}}
\end{equation}
where $\times$ indicates interaction. Socio-technical alignment $STA$ is a continuous value in $(0,2]$. $STA < 1$ indicates a well-aligned, resilient system that reduces risk, $STA = 1$ indicates sufficient alignment with no impact on the risk, and $STA > 1$ indicates misalignment that amplifies risk. Alignment means subsystem attributes match, e.g., a well-aligned system has an explainable AI system, transparent communication of tasks and limits, trained operators, and a strong safety culture. A misaligned system may have reliable monitoring of AI systems but untrained operators and an organization that prioritizes speed over safety. To estimate $STA$, we propose measuring key socio-technical attributes, outlined in Section~\ref{sec:4}, via subsystem proxies, expressed as:
\begin{equation} \label{eq:sta}
{STA = f(STA_{tech},STA_{human},STA_{org})}
\end{equation}
where $STA_{tech}$ quantifies technical alignment, such as the AI system transparency including whether and how the AI system or its developers communicate the system's task, internal operation, operational environment, and limitations. $STA_{human}$ measures human alignment, including the operator's competence, relevant prior knowledge, and trust in the system. Finally, $STA_{org}$ assesses organizational alignment, for instance through the strength of safety culture and the quality of training protocols. To achieve $STA<1$, all three attributes must be aligned. A weakness in any one dimension, e.g., strong technical transparency but insufficient operator training and competence, negatively impacts the overall socio-technical alignment. The assessment of these attributes depends on the context but requires a mixed-methods approach combining quantitative metrics and qualitative evaluations.

\section{Case study: AI-enabled $LH_2$ bunkering system}
\begin{figure}
\begin{center}
\includegraphics[width=7cm]{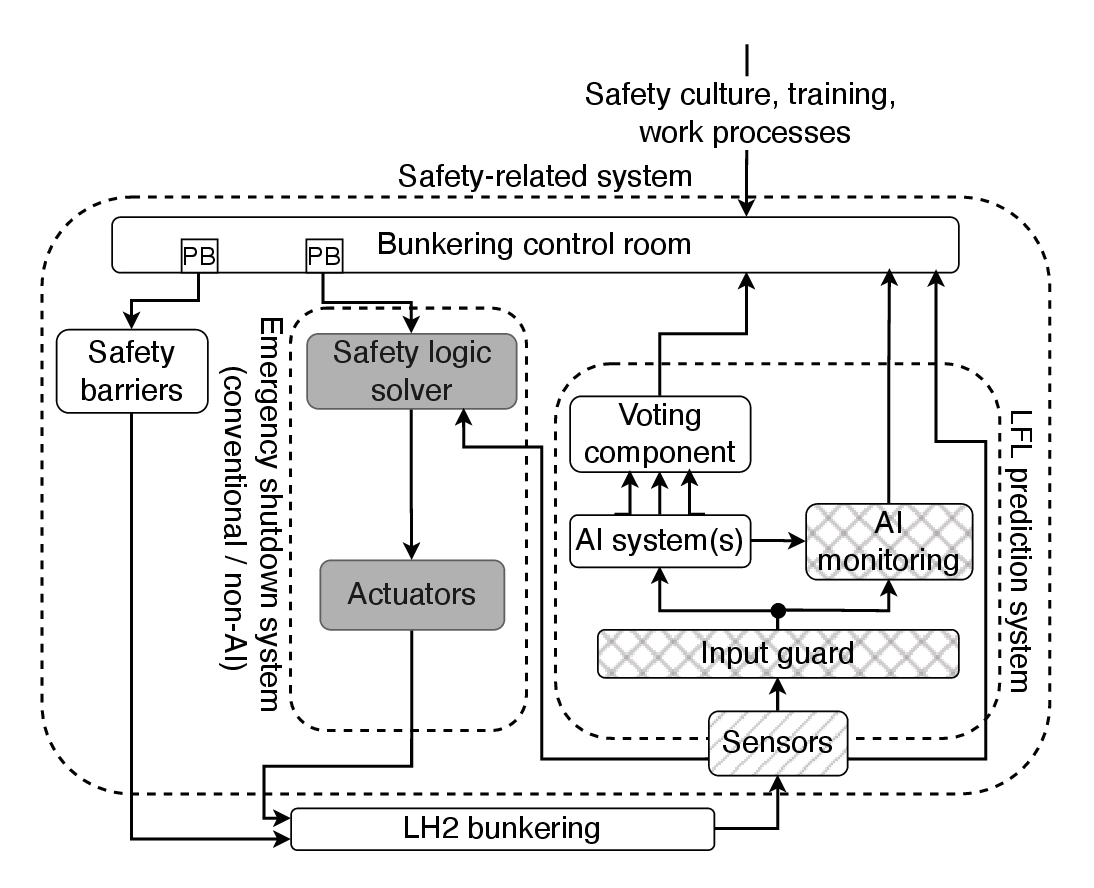}
\caption{Safeguarded system design of an AI-enabled $LH_2$ bunkering system based on \cite{flehmigImplementingArtificialIntelligence2024a}. The cross-hatched components are the new components.} 
\label{fig:robust}
\end{center}
\end{figure}
\begin{figure}
\begin{center}
\includegraphics[width=8.4cm]{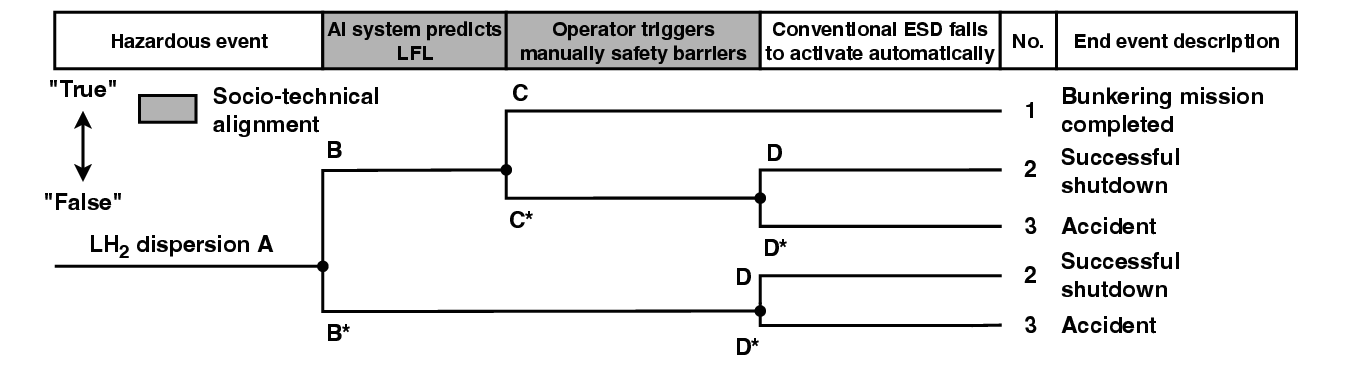}
\caption{Event tree for $LH_2$ dispersion within an AI-enabled $LH_2$ bunkering system.} 
\label{fig:ETA}
\end{center}
\end{figure}

Finally, we conduct a validation of the proposed socio-technical alignment variable. We evaluate the risk for an AI-enabled safety-critical system both before and after a redesign, demonstrating how the novel risk expression captures socio-technical attributes and their alignment compared to the traditional formulation.
We apply this to an AI-enabled $LH_2$ bunkering system, where an AI system serves as a decision-support tool for human operators in the bunkering control room. The initial design of a non-AI $LH_2$ bunkering system with a bunkering control room, an emergency shutdown (ESD) system and other system-specific design choices is based on \citep{tamburiniAnalysisSystemResilience2025, europeanmaritimesafetyagencyHazardIdentificationGeneric2024, maritimetechnologiesforumGuidelinesDevelopmentLiquefied2024}. A potential AI system for $LH_2$ bunkering systems was introduced in \cite{alfariziAccidentPreventionLiquid2023}. The authors created an AI system that predicts whether the lower flammability limit (LFL) of the hydrogen is exceeded in 200 seconds. For this work, we incorporate this AI system in the $LH_2$ bunkering system. The incorporated AI system enables operators to trigger safety barriers that can prevent an emergency shutdown that is automatically triggered when the LFL is exceeded. This system forms a socio-technical system: bunkering control room operators are human operators, the AI system with its sensors for perception are technical components, and bunkering procedures and operating organization represent the organizational structure. This system also qualifies as an AI-enabled, safety-critical system because of the integration of the AI system into the broader $LH_2$ bunkering system, and failures of this system could result in significant harm to people or the environment.\\
We analyze two distinct system designs. The first is a naive implementation without specific safeguards except a voting mechanism, as seen in Figure~\ref{fig:robust} excluding the cross-hatched components. The second is a safeguarded system design, Figure~\ref{fig:robust}, designed according to \cite{flehmigImplementingArtificialIntelligence2024a} and aligned with the functional safety guidance for AI systems in \cite{iso/iectr5469:2024ArtificialIntelligenceFunctional2024}, as intended for safety-critical applications. This design essentially introduces two new system components dedicated to monitoring the AI system and online validation of input data. To compare the traditional risk equation with our proposed socio-technical extended version, we consider a specific hazardous event in $LH_2$ bunkering: hydrogen dispersion. Figure \ref{fig:ETA} presents the corresponding event tree for an AI-enabled $LH_2$ bunkering system and Figure \ref{fig:robust} shows the different system designs, naive and safeguarded. 
The traditional risk equation evaluates scenario likelihoods and consequences, but socio-technical attributes are easily overlooked. It emphasizes technical aspects and component failures over interaction failures, typically suggests a net risk reduction when components such as AI monitoring and input guards are added. These components reduce the number of events where the AI fails to predict LFL (\(B^*\downarrow\)), thereby decreasing accidents and successful shutdowns resulting from missed detection (\(A \cap B^* \downarrow \cap (D \cup D^*)\)). Additionally, they reduce accidents and successful shutdowns in other branches (\(A \cap B \uparrow \cap C^* \downarrow \cap (D \cup D^*)\)) while increasing the number of completed bunkering missions (\(A \cap B \uparrow \cap C \uparrow\)), leading to an overall risk reduction. In contrast, our proposed equation explicitly incorporates socio-technical attributes and their alignment. As highlighted in gray in Figure \ref{fig:ETA}, these events are shaped by socio-technical attributes, and adding monitoring or input guards does not guarantee risk reduction. Such components cannot be evaluated in isolation but must be considered within their socio-technical context. They increase system complexity and can amplify risk by affecting the probabilities of pivotal events $B$, $B^*$, $C$, and $C^*$. For example, poorly explained monitoring degrades technical alignment (\(STA_{\text{tech}}\)); greater task complexity demands higher operator competence, harming human alignment (\(STA_{\text{human}}\)); and conflicts between AI, monitoring, and human expertise require updated processes and training, impacting organizational alignment (\(STA_{\text{org}}\)). These factors influence $B$, $B^*$, $C$, and $C^*$ in ways that may either reduce or amplify risk, depending on deeper assessment. Ultimately, simply adding technical safety layers without considering interactions among operators, technical systems, and organizational structures does not necessarily lower risk. Without sufficient attention to socio-technical attributes, such additions may counterintuitively increase overall system risk. This underscores the need to explicitly account for socio-technical attributes, a dimension where the traditional risk equation falls short.

\section{Discussion}

This work identifies a gap in the risk evaluation: the lack of explicitly integrating socio-technical attributes, despite AI-enabled systems can be qualified as socio-technical systems. To address this, we propose a novel variable to the traditional risk expression, designed to explicitly incorporate socio-technical attributes through subsystem proxies.\\
First, while AI-enabled systems are acknowledged as complex \citep{dobbeSystemSafetyArtificial2022a, hendrycksIntroductionAISafety2024a}, their explicit characterization as socio-technical systems, derived from a comparison of key attributes, remains an interpretative step. This assumption is central to our analysis, but we acknowledge that other perspectives may differ.\\
Second, the socio-technical system perspective may also be applicable to systems employing advanced non-AI processing algorithms, as noted by \cite{levesonEngineeringSaferWorld2012} regarding modern complex systems, however key differences between AI systems and conventional advanced algorithms significantly impact socio-technical attributes and can lead to more severe risk. Advanced algorithms operate on human-programmed rules, whereas AI systems autonomously derive rules through optimization of an objective function, often yielding intractable and uninterpretable outcomes; the "black box" problem. This opacity affects critical socio-technical attributes such as complexity and interdependence. Moreover, AI systems can exhibit emergent capabilities beyond those of traditional algorithms. For instance, large language models initially designed for text generation were later found capable of providing instructions for synthesizing biological weapons. These distinctions do not argue against applying the socio-technical alignment variable to systems with advanced algorithms. Instead, they underscore why such alignment is explicitly necessary for AI-enabled systems.\\
Third, the scope of our literature review, which prioritized established industry standards \citep{iec310102019Risk2020, InformationTechnologyArtificial2024, RiskManagementGuidelines2018} and key supplemental research \citep{koesslerRiskAssessmentAGI2023, hendrycksIntroductionAISafety2024a}, was targeted rather than exhaustive. It was designed to examine practices directly relevant to industrial implementation. While sufficient to identify a gap in current risk evaluation techniques, a more comprehensive review would be required to make a definitive general statement.\\
Fourth, the identification of this gap that is the absence of an explicit variable for socio-technical alignment in common risk equations, justifies the introduction of our proposed variable. Given the distinct safety dynamics inherent to socio-technical systems, its inclusion is a meaningful step towards a more holistic AI risk evaluation. It is also needed due to empirical shortcomings in existing risk assessment when it comes to socio-technical attributes of AI-enabled systems \citep{rajiConcreteProblemsAI2024}. However, we recognize that socio-technical attributes can be captured in the traditional risk equation but rather as latent variables, and thus may be neglected.\\
Fifth, the proposed socio-technical alignment $STA$ variable is intentionally simplified, using subsystem proxies to ensure initial tractability and usability. This foundational idea is not intended to be exhaustive but to establish a basis for future refinement and deeper investigation into measuring or determining socio-technical attributes. Key challenges include the availability of relevant data and the domain expertise required to meaningfully assess these attributes, as well as the difficulty of determining whether to weight different dimensions equally or to prioritize certain dimensions over others. Finally, the case study demonstrated the potential relevance of the variable within a realistic AI-enabled safety-critical context. While it served to highlight the differences between traditional and extended risk evaluation, the demonstration was not sufficient for formal validation, which would require more rigorous, empirical testing.
\section{Conclusion} 
This work established AI-enabled systems as socio-techni-cal systems, demonstrated that existing risk evaluation methods inadequately account for socio-technical attributes, and introduced a novel variable, named the socio-technical alignment, to explicitly incorporate these attributes into the traditional risk expression. The proposed risk equation may foster increased awareness of socio-technical attributes during risk assessment of AI-enabled systems, thereby enhancing the system design process and mitigating risks arising from these attributes. A primary direction for future work is to test this variable by developing and validating suitable proxies for its core attributes, enabling its semi-quantitative or quantitative assessment in practice.

\begin{ack}
This work was carried out as a part of SUBPRO-Zero, a Research Centre at NTNU. The authors gratefully acknowledge the project support from SUBPRO-Zero, which is financed by major industry partners and NTNU. We also thank Abhishek Subedi for helpful feedback on socio-technical systems and the risk assessment process.
\end{ack}

\section*{DECLARATION OF GENERATIVE AI AND AI-ASSISTED TECHNOLOGIES IN THE WRITING PROCESS}
During the preparation of this work the author(s) used DeepSeek in order to help with editing. After using this tool, the author(s) reviewed and edited the content as needed and take(s) full responsibility for the content of the publication.

\bibliography{ifacconf}           

@book{hendrycksIntroductionAISafety2024a,
  title = {Introduction to AI Safety, Ethics, and Society},
  author = {Hendrycks, Dan},
  year = {2024},
  publisher = {Taylor \& Francis CRC Press [Imprint]},
  location = {Erscheinungsort nicht ermittelbar},
  isbn = {978-1-040-26117-0 978-1-003-53033-6 978-1-032-86992-6 978-1-040-26116-3},
  langid = {und},
  pagetotal = {1}
}

@misc{InformationTechnologyArtificial2024,
  type = {{{NAT}}},
  title = {Information Technology - {{Artificial}} Intelligence - {{Guidance}} on Risk Management},
  author = {{ISO/IEC 23894}},
  year = 2024,
  month = mar,
  urldate = {2025-11-20}
}

@misc{koesslerRiskAssessmentAGI2023,
  title = {Risk Assessment at {{AGI}} Companies: {{A}} Review of Popular Risk Assessment Techniques from Other Safety-Critical Industries},
  shorttitle = {Risk Assessment at {{AGI}} Companies},
  author = {Koessler, Leonie and Schuett, Jonas},
  year = 2023,
  publisher = {arXiv},
  doi = {10.48550/ARXIV.2307.08823},
  urldate = {2025-12-01},
  copyright = {Creative Commons Attribution 4.0 International}
}

@book{levesonEngineeringSaferWorld2012,
  title = {Engineering a {{Safer World}}: {{Systems Thinking Applied}} to {{Safety}}},
  shorttitle = {Engineering a {{Safer World}}},
  author = {Leveson, Nancy G.},
  year = 2012,
  series = {Engineering {{Systems}}},
  publisher = {The MIT Press},
  address = {Cambridge},
  isbn = {978-0-262-29824-7 978-0-262-01662-9},
  langid = {english}
}

@inproceedings{lootzRiskManagementUncertainty2025,
  title = {Risk {{Management}} and {{Uncertainty}} of {{Artificial Intelligence}} in a {{High Hazard Industry}}},
  booktitle = {35th {{European Safety}} and {{Reliability Conference}} ({{ESREL}} 2025) and the 33rd {{Society}} for {{Risk Analysis Europe Conference}} ({{SRA-E}} 2025)},
  author = {Lootz, Elisabeth and Vestly Bergh, Linn Iren and Heide, Bj{\o}rnar},
  year = 2025,
  pages = {1420--1427},
  publisher = {Research Publishing Services},
  address = {Singapore EXPO, Singapore},
  doi = {10.3850/978-981-94-3281-3\_ESREL-SRA-E2025-P2957-cd},
  urldate = {2025-10-24},
  isbn = {978-981-94-3281-3},
  langid = {english}
}

@incollection{nagbolDesigningRiskAssessment2021,
  title = {Designing a {{Risk Assessment Tool}} for {{Artificial Intelligence Systems}}},
  booktitle = {The {{Next Wave}} of {{Sociotechnical Design}}},
  author = {Nagb{\o}l, Per R{\aa}dberg and M{\"u}ller, Oliver and Krancher, Oliver},
  editor = {Chandra Kruse, Leona and Seidel, Stefan and Hausvik, Geir Inge},
  year = 2021,
  volume = {12807},
  pages = {328--339},
  publisher = {Springer International Publishing},
  address = {Cham},
  doi = {10.1007/978-3-030-82405-1\_32},
  urldate = {2025-11-21},
  isbn = {978-3-030-82404-4 978-3-030-82405-1},
  langid = {english}
}

@article{novelliAIRiskAssessment2024,
  title = {{{AI Risk Assessment}}: {{A Scenario-Based}}, {{Proportional Methodology}} for the {{AI Act}}},
  shorttitle = {{{AI Risk Assessment}}},
  author = {Novelli, Claudio and Casolari, Federico and Rotolo, Antonino and Taddeo, Mariarosaria and Floridi, Luciano},
  year = 2024,
  month = may,
  journal = {Digital Society},
  volume = {3},
  number = {1},
  pages = {13},
  issn = {2731-4650, 2731-4669},
  doi = {10.1007/s44206-024-00095-1},
  urldate = {2025-11-21},
  langid = {english}
}

@misc{RiskManagementGuidelines2018,
  author = {{ISO 31000}},
  title = {Risk Management --- {{Guidelines}}},
  year = 2018,
  month = feb
}

@misc{camposFrontierAIRisk2025,
  title = {A {{Frontier AI Risk Management Framework}}: {{Bridging}} the {{Gap Between Current AI Practices}} and {{Established Risk Management}}},
  shorttitle = {A {{Frontier AI Risk Management Framework}}},
  author = {Campos, Simeon and Papadatos, Henry and Roger, Fabien and Touzet, Chloé and Quarks, Otter and Murray, Malcolm},
  year = 2025,
  doi = {10.48550/ARXIV.2502.06656},
  pubstate = {prepublished},
  version = {3}
}

@misc{euaiactEuropeanUnionArtificial2024,
  title = {European {{Union Artificial Intelligence Act}}},
  author = {{EU AI Act}},
  year = 2024,
  publisher = {European Union},
  langid = {english}
}

@misc{iso/iectr5469:2024ArtificialIntelligenceFunctional2024,
  title = {Artificial Intelligence — {{Functional}} Safety and {{AI}} Systems},
  author = {{ISO/IEC TR 5469}},
  year = 2024,
  langid = {english},
}

@book{rausandRiskAssessmentTheory2020,
  title = {Risk {{Assessment}}: {{Theory}}, {{Methods}}, and {{Applications}}},
  shorttitle = {Risk {{Assessment}}},
  author = {Rausand, Marvin and Haugen, Stein},
  year = 2020,
  edition = {1},
  publisher = {Wiley},
  doi = {10.1002/9781119377351},
  isbn = {978-1-119-37723-8 978-1-119-37735-1},
  langid = {english}
}

@book{zareiSafetyCausationAnalysis2024,
  title = {Safety {{Causation Analysis}} in {{Sociotechnical Systems}}: {{Advanced Models}} and {{Techniques}}},
  shorttitle = {Safety {{Causation Analysis}} in {{Sociotechnical Systems}}},
  editor = {Zarei, Esmaeil},
  year = 2024,
  series = {Studies in {{Systems}}, {{Decision}} and {{Control}}},
  volume = {541},
  publisher = {Springer Nature Switzerland},
  location = {Cham},
  doi = {10.1007/978-3-031-62470-4},
  isbn = {978-3-031-62469-8 978-3-031-62470-4},
  langid = {english}
}

@misc{detnorskeveritasAssuranceAIenabledSystems2023,
  type = {Recommended practice},
  title = {Assurance of {{AI-enabled}} Systems},
  author = {{Det Norske Veritas}},
  year = 2023,
  number = {DNV-RP-0671},
  institution = {Det Norske Veritas}
}

@misc{europeanmaritimesafetyagencyHazardIdentificationGeneric2024,
  title = {Hazard Identification of Generic Hydrogen Fuel Systems},
  author = {{European Maritime Safety Agency}},
  year = 2024,
  location = {Lisbon},
}

@inproceedings{flehmigImplementingArtificialIntelligence2024a,
  title = {Implementing {{Artificial Intelligence}} in {{Safety-Critical Systems}} during {{Operation}}: {{Challenges}} and {{Extended Framework}} for a {{Quality Assurance Process}}},
  shorttitle = {Implementing {{Artificial Intelligence}} in {{Safety-Critical Systems}} during {{Operation}}},
  booktitle = {{{IECON}} 2024 - 50th {{Annual Conference}} of the {{IEEE Industrial Electronics Society}}},
  author = {Flehmig, Niclas and Lundteigen, Mary Ann and Yin, Shen},
  year = 2024,
  pages = {1--8},
  publisher = {IEEE},
  location = {Chicago, IL, USA},
  doi = {10.1109/IECON55916.2024.10906021},
  eventtitle = {{{IECON}} 2024 - 50th {{Annual Conference}} of the {{IEEE Industrial Electronics Society}}},
  isbn = {978-1-6654-6454-3}
}

@misc{maritimetechnologiesforumGuidelinesDevelopmentLiquefied2024,
  title = {Guidelines for the Development of Liquefied Hydrogen Bunkering Systems and Procedures},
  author = {{Maritime Technologies Forum}},
  year = 2024,
  institution = {Det Norske Veritas},
}

@article{alfariziAccidentPreventionLiquid2023,
  title = {Towards Accident Prevention on Liquid Hydrogen: {{A}} Data-Driven Approach for Releases Prediction},
  shorttitle = {Towards Accident Prevention on Liquid Hydrogen},
  author = {Alfarizi, Muhammad Gibran and Ustolin, Federico and Vatn, Jørn and Yin, Shen and Paltrinieri, Nicola},
  year = 2023,
  journal = {Reliability Engineering \& System Safety},
  shortjournal = {Reliability Engineering \& System Safety},
  volume = {236},
  pages = {109276},
  issn = {09518320},
  doi = {10.1016/j.ress.2023.109276},
  langid = {english}
}

@article{tamburiniAnalysisSystemResilience2025,
  title = {Analysis of System Resilience in Escalation Scenarios Involving {{LH2}} Bunkering Operations},
  author = {Tamburini, Federica and Iaiani, Matteo and Cozzani, Valerio},
  year = 2025,
  journal = {Reliability Engineering \& System Safety},
  shortjournal = {Reliability Engineering \& System Safety},
  volume = {257},
  pages = {110816},
  issn = {09518320},
  doi = {10.1016/j.ress.2025.110816},
  langid = {english}
}

@misc{iec310102019Risk2020,
  type = {NAT},
  title = {{{Risk}} Management - {{Risk}} Assessment Techniques},
  author = {{IEC 31010}},
  year = 2020
}

@inproceedings{dobbeSystemSafetyArtificial2022a,
  title = {System {{Safety}} and {{Artificial Intelligence}}},
  booktitle = {2022 {{ACM Conference}} on {{Fairness Accountability}} and {{Transparency}}},
  author = {Dobbe, Roel},
  year = {2022},
  pages = {1584--1584},
  publisher = {ACM},
  location = {Seoul Republic of Korea},
  doi = {10.1145/3531146.3533215},
  eventtitle = {{{FAccT}} '22: 2022 {{ACM Conference}} on {{Fairness}}, {{Accountability}}, and {{Transparency}}},
  isbn = {978-1-4503-9352-2},
  langid = {english}
}

@article{kaplanQuantitativeDefinitionRisk1981,
  title = {On {{The Quantitative Definition}} of {{Risk}}},
  author = {Kaplan, Stanley and Garrick, B. John},
  year = 1981,
  journal = {Risk Analysis},
  shortjournal = {Risk Analysis},
  volume = {1},
  number = {1},
  pages = {11--27},
  issn = {0272-4332, 1539-6924},
  doi = {10.1111/j.1539-6924.1981.tb01350.x},
  langid = {english}
}

@article{rasmussenRiskManagementDynamic1997,
  title = {Risk Management in a Dynamic Society: A Modelling Problem},
  shorttitle = {Risk Management in a Dynamic Society},
  author = {Rasmussen, Jens},
  year = 1997,
  journal = {Safety Science},
  shortjournal = {Safety Science},
  volume = {27},
  number = {2--3},
  pages = {183--213},
  issn = {09257535},
  doi = {10.1016/S0925-7535(97)00052-0},
  langid = {english}
}

@misc{rajiConcreteProblemsAI2024,
  title = {Concrete {{Problems}} in {{AI Safety}}, {{Revisited}}},
  author = {Raji, Inioluwa Deborah and Dobbe, Roel},
  year = {2024},
  doi = {10.48550/ARXIV.2401.10899},
  pubstate = {prepublished},
  version = {1}
}

@misc{AI_managementSystem,
  type = {NAT},
  title = {Information Technology - {{Artificial}} Intelligence - {{Mangement}} System},
  author = {{ISO/IEC 42001}},
  year = {2023},
  version = {1}
}
\end{document}